\documentclass[aps,prd,superscriptaddress,nofootinbib]{revtex4-2}

\usepackage[utf8]{inputenc}
\usepackage[T1]{fontenc}
\usepackage{amsmath}
\usepackage{amssymb}
\usepackage{amsfonts}
\usepackage{graphicx}  
\usepackage{hyperref}  
\usepackage{slashed}   
\usepackage{booktabs}  
\usepackage{microtype} 

\hypersetup{
    colorlinks=true,
    linkcolor=blue,
    filecolor=magenta,      
    urlcolor=blue,
    citecolor=blue,
}

\begin{document}

\title{\textbf{The Dirac-Majorana Confusion Theorem in the Presence of Flavor-Changing Neutral Currents: A CP-Filter Mechanism}}

\author{David Delepine}
\email{delepine@ugto.mx}
\affiliation{Divisi\'on de Ciencias e Ingenier\'ias, Universidad de Guanajuato, C.P. 37150, Le\'on, Guanajuato, M\'exico.}

\author{A. Yebra}
\email{azarael@fisica.ugto.mx}
\affiliation{Divisi\'on de Ciencias e Ingenier\'ias, Universidad de Guanajuato, C.P. 37150, Le\'on, Guanajuato, M\'exico.}

\date{\today}

\begin{abstract}
The ``Practical Dirac-Majorana Confusion Theorem'' (PDMCT) asserts that phenomenological differences between Dirac and Majorana neutrinos are kinematically suppressed by $(m_\nu/E)^2$ in lepton-number-conserving processes. The PDMCT relies on standard left-handed neutrino interactions, as in the SM, in which case Dirac and Majorana neutrinos are experimentally indistinguishable in lepton-number-conserving processes. Our scenario explicitly goes beyond these assumptions by introducing a neutral vector boson $Z^{'}$ with CP-violating, flavor-changing neutral current (FCNC) couplings, which generate observable Dirac–Majorana differences while remaining fully consistent with the PDMCT. In strict accordance with the PDMCT, Fermi-Dirac statistics dictate that the flavor-diagonal vector current identically vanishes for Majorana neutrinos.For non-diagonal transitions, the Majorana condition strictly forbids the real (CP-conserving) component of the vector interaction, leaving only a imaginary coupling. As a consequence, the observable difference in inclusive scattering cross-sections between Dirac and Majorana neutrinos is driven entirely by the FCNC CP-violating couplings. We apply these results to Coherent Elastic Neutrino-Nucleus Scattering (CE$\nu$NS), showing that for spin-zero targets, the distinguishability of the neutrino's nature is  determined by the CP structure of the new interaction.
\end{abstract}

\maketitle

\section{Introduction}

The determination of the neutrino nature —whether it is a Dirac fermion ($\nu \neq \bar{\nu}$) or a Majorana fermion ($\nu = \nu^c$)—remains one of the most important open questions in Physics. While the observation of Neutrinoless Double Beta Decay ($0\nu\beta\beta$) \cite{Furry1939,Schechter1982,Rodejohann2011,Dolinski2019,KamLAND-Zen2023} would unambiguously confirm the Majorana nature by proving the violation of total lepton number ($L$), constraints from current experiments remain inconclusive.

In the absence of $0\nu\beta\beta$ detection, it is theoretically possible to distinguish Dirac and Majorana neutrinos through scattering experiments, such as elastic neutrino-electron scattering ($\nu e^- \to \nu e^-$) or Coherent Elastic Neutrino-Nucleus Scattering (CE$\nu$NS). However, the \textit{Practical Dirac-Majorana Confusion Theorem} (PDMCT), established by Kayser and Shrock \cite{Kayser:1982jw,Kayser:1982wd,Rosen:1982pj}, states that for any lepton number conserving process, the difference between Dirac and Majorana cross-sections is proportional to the square of the neutrino mass relative to the energy scale:
\begin{equation}
    \sigma_{\text{Majorana}} - \sigma_{\text{Dirac}} \propto \left( \frac{m_\nu}{E_\nu} \right)^2.
\end{equation}

Given the sub-eV scale of neutrino masses and the MeV-GeV energies of typical experiments, this difference is of the order $\lesssim \mathcal{O}(10^{-14})$, rendering the two cases experimentally indistinguishable within the Standard Model (SM).

 For a Majorana field, Fermi-Dirac statistics and the charge conjugation properties  dictate that the diagonal vectorial bilinear identically vanishes ($\bar{\nu}_\alpha\gamma^\mu \nu_\alpha \equiv 0$). Consequently, even in the presence of New Physics (NP) with complex phases, the diagonal vectorial interaction cannot generate a distinguishing signal.

Recent literature \cite{Bigaran2025, Marquez2024,Marquez2024_PRD, Akhmedov2024} has  clarified and reaffirmed the absolute robustness of the PDMCT, resolving recent community confusion by demonstrating that the theorem holds  when the underlying parameter spaces are properly evaluated. In particular, these works explicitly emphasize that the PDMCT is derived assuming SM-like, purely left-handed neutrino interactions and flavor-conserving neutral currents, and that any observable Dirac–Majorana difference requires going beyond these hypotheses.In complete agreement with these works, we explore a scenario that  arises from the  application of the theorem to non-diagonal interactions for Majorana neutrino: the existence of New Physics mediated by a neutral vector boson $Z'$ with \textit{complex}, Flavor-Changing Neutral Current (FCNC) couplings that violate CP symmetry.As the PDMCT  forbids CP-conserving vector FCNCs for Majorana fermions, it creates a phenomenological difference between Dirac and Majorana states that relies on the CP-violating phase. 
If we define the effective vector FCNC coupling as $g_V^{\alpha\beta} = |g| e^{i\phi}$:

\begin{itemize}

\item A \textbf{Dirac neutrino} interacts via the full modulus $|g_V^{\alpha\beta}|^2$, and its inclusive cross-section is dominated by the Standard Model interference in the diagonal channel.
\item A \textbf{Majorana neutrino}, as $\phi$ is fixed to $\pm \pi/2$, identically decouples from the diagonal SM vector interaction as mandated by the PDMCT, but couples to the new FCNC transitions \textit{exclusively} through a purely imaginary, CP-violating vector component.
\end{itemize}
This leads to a distinct phenomenological signature for inclusive scattering measurements where the outgoing neutrino flavor is not detected. 

In Section II, we develop the formalism for the effective Majorana flavor-changing couplings. Section III presents the calculation of the inclusive differential cross-sections for $\nu-e$ scattering and CE$\nu$NS. Finally, in Section IV, we discuss the experimental implications for current and future detectors such as COHERENT and DUNE.

\section{FORMALISM: FLAVOR-CHANGING VECTOR INTERACTIONS AND HERMITICITY}
\label{sec:formalism}

Motivated by recent discussions reaffirming the absolute robustness of the
Practical Dirac--Majorana Confusion Theorem (PDMCT)%
~\cite{Kayser:1982wd,Kayser:1982jw,Rosen:1982pj,Bigaran2025, Marquez2024,Marquez2024_PRD,Akhmedov2024},
we focus specifically on flavor-changing neutral currents (FCNC).

Within the Standard Model, and more generally in any theory satisfying the
assumptions of the PDMCT, neutrinos participate only in purely left-handed
weak interactions and the neutral current is flavor-diagonal.\footnote{In
this situation, Dirac and Majorana neutrinos yield identical neutral-current
cross sections up to corrections of order $(m_\nu/E_\nu)^2$, as discussed in
Refs.~\cite{Kayser:1982wd,Kayser:1982jw,Rosen:1982pj,Bigaran2025, Marquez2024,Marquez2024_PRD,Akhmedov2024}.}
In this limit, neutral-current processes such as elastic scattering or CE$\nu$NS
cannot distinguish the neutrino nature, in full agreement with the PDMCT.

In what follows we introduce \emph{additional} interactions mediated by a new
neutral vector boson $Z'$ that couples to neutrinos through complex,
flavor-changing neutral currents. The ``CP-filter'' effect that we identify
in this work is therefore a property of these extra FCNC couplings of the
$Z'$ boson, and not of the Standard Model neutral current itself.
Let us construct the most general effective flavor-changing vector interaction.

We define a coupling $c_{\alpha\beta}$ for the transition between flavors $\alpha$ and $\beta$. The interaction can be written as a single term:

\begin{equation}\mathcal{L}_{\text{int}} \supset - Z'_{\mu} \left( c_{\alpha\beta} \overline{\nu}_{\alpha} \gamma^\mu \nu_{\beta} \right).\label{eq:lagrangian_fcnc}
\end{equation}

A fundamental requirement of any physical theory is that the Lagrangian must be Hermitian ($\mathcal{L} = \mathcal{L}^\dagger$). The phenomenological consequences of this requirement depend intrinsically on the fundamental nature of the neutrino field.

\subsection{Dirac Neutrinos}
If neutrinos are Dirac fermions, $\nu_\alpha$ and $\nu_\beta$ are distinct from their antiparticles. The bilinear $\overline{\nu}_\alpha \gamma^\mu \nu_\beta$ and its conjugate $\overline{\nu}_\beta \gamma^\mu \nu_\alpha$ are independent operators mediating distinct physical processes ($\nu_\beta \to \nu_\alpha$ and $\nu_\alpha \to \nu_\beta$). To ensure Hermiticity, the Lagrangian must explicitly contain both the term and its Hermitian conjugate:

\begin{equation}\mathcal{L}_{\text{int}}^{\text{Dirac}} \supset - Z'_{\mu} \left( g_{\alpha\beta} \overline{\nu}_{\alpha} \gamma^\mu \nu_{\beta} + g_{\alpha\beta}^* \overline{\nu}_{\beta} \gamma^\mu \nu_{\alpha }\right),
\end{equation}
where we identify the coupling as a general complex number $c_{\alpha\beta} = g_{\alpha\beta}$. For a scattering experiment measuring the transition $\nu_\alpha \to \nu_\beta$, the cross-section is proportional to the full magnitude of the coupling $|g_{\alpha\beta}|^2$, making a Dirac neutrino sensitive to both the CP-conserving (real) and CP-violating (imaginary) components.\subsection{Majorana Neutrinos and the CP-Filter Mechanism}If neutrinos are Majorana fermions, they satisfy the self-conjugation condition $\nu = \nu^c = C\overline{\nu}^T$, where $C$ is the charge conjugation matrix. This imposes strict algebraic constraints on the bilinear covariants. Following Denner's rules for fermion-number-violating interactions \cite{Denner1992}, we apply the properties of the charge conjugation matrix ($C^T = -C$ and $C \gamma^\mu C^{-1} = -(\gamma^\mu)^T$) alongside the anticommutation relations of fermion fields (Grassmann variables).Transposing the vector bilinear yields:

\begin{equation}(\overline{\nu}_\beta \gamma^\mu \nu_{\alpha})^T = (\nu_\beta^T C \gamma^\mu \nu_\alpha)^T = - \nu_\alpha^T (\gamma^\mu)^T C^T \nu_\beta = - \nu_\alpha^T C \gamma^\mu \nu_\beta = - \overline{\nu}_\alpha \gamma^\mu \nu_{\beta}.\label{eq:majorana_identity}
\end{equation}

The minus sign arises strictly from Fermi-Dirac statistics \cite{Denner1992}.
So, our $\mathcal{L}_{\text{int}}$ is given by:
\begin{equation}[\mathcal{L}_{\text{int}}]^\dagger = - Z'_\mu c_{\alpha\beta}^* (\overline{\nu}_\alpha \gamma^\mu \nu_\beta)^\dagger = - Z'_\mu c_{\alpha\beta}^* (\overline{\nu}_\beta \gamma^\mu \nu_\alpha) = + Z'_\mu c_{\alpha\beta}^* (\overline{\nu}_\alpha \gamma^\mu \nu_\beta).
\end{equation}

For the Majorana Lagrangian to be physically valid and Hermitian ($\mathcal{L}_{\text{int}} = \mathcal{L}_{\text{int}}^\dagger$), we must enforce the condition:\begin{equation}-c_{\alpha\beta} = c_{\alpha\beta}^* \implies \operatorname{Re}(c_{\alpha\beta}) = 0,\end{equation}
The mathematical structure of Majorana fermion acts as an absolute filter, forbidding real (CP-conserving) vector FCNC couplings. 
We can therefore define the effective Majorana coupling as $c_{\alpha\beta} = i (g_{\alpha\beta})$ where $g_{\alpha\beta}$ is real.

\section{Scattering Cross-Sections and Phenomenological Consequences}

We now apply the effective coupling formalism to two distinct scattering processes: elastic neutrino-electron scattering (ES) and Coherent Elastic Neutrino-Nucleus Scattering (CE$\nu$NS). We assume the limit $E_\nu \gg m_\nu$, where the kinematic mass suppression from the standard confusion theorem is negligible ($\lesssim \mathcal{O}(10^{-14})$), allowing us to isolate the effects induced purely by the CP-violating phases.

\subsection{Elastic Neutrino-Electron Scattering ($\nu_\alpha e^- \to \nu_x e^-$)}

In standard neutrino-electron scattering experiments, the initial neutrino flavor $\alpha$ is typically known (e.g., $\nu_e$ from a reactor or $\nu_\mu$ from an accelerator beam). However, the outgoing neutrino is not detected; only the recoiling electron is measured. Consequently, the observed differential cross-section is an inclusive sum over all possible final neutrino flavors $\beta$:
\begin{equation}
\frac{d\sigma_{\text{obs}}}{dT} = \sum_{\beta=e,\mu,\tau} \frac{d\sigma_{\alpha\beta}}{dT} \text{.}
\end{equation}
Experimentally, the energy resolution of the detectors in these scattering processes cannot kinematically resolve the individual outgoing neutrino mass eigenstates ($E_\nu \gg m_i$). Because these final states are kinematically degenerate and unobserved, we must perform an inclusive sum over all possible final states. By virtue of the unitarity of the PMNS mixing matrix ($\sum_{i} U_{\beta i} U_{\gamma i}^* = \delta_{\beta \gamma}$), the incoherent sum over the mass eigenstates is  identical to the sum over the flavor eigenstates. Therefore, it is   more transparent for evaluating New Physics couplings—to perform the cross-section sum directly in the flavor basis.

Since the final-state neutrinos are not detected, one must sum over all possible outgoing states. Because this is an inclusive cross-section where the specific final mass states are unresolved, we can rigorously perform this sum directly in the flavor basis without needing to rotate to the mass basis.

The differential cross-section with respect to the electron recoil energy $T$ for a specific transition from flavor $\alpha$ to $\beta$ is given by \cite{Kayser:1982wd,Kayser:1982jw}:
\begin{equation}
\frac{d\sigma_{\alpha\beta}}{dT} = \frac{2 G_F^2 m_e}{\pi} \left[ |g_L^{\alpha\beta}|^2 + |g_R^{\alpha\beta}|^2 \left(1 - \frac{T}{E_\nu}\right)^2 - \frac{m_e T}{E_\nu^2} \operatorname{Re}(g_L^{\alpha\beta} g_R^{\alpha\beta *}) \right] \text{,}
\end{equation}
where $g_L^{\alpha\beta}$ and $g_R^{\alpha\beta}$ are the effective chiral couplings.

If we restrict the effective couplings  $g_{L.R}^{\alpha \beta}$ to their SM values and to left-handed interactions, our expression reproduces the well-known result that the $\nu$–e cross section is identical for Dirac and Majorana neutrinos, in agreement with the analysis of Ref. \cite{Marquez2024} , see their eqs. (2.1)–(2.6) and Section 2.1. In this limit, the parameters characterizing the cross section take the same values for both neutrino natures, and no Dirac–Majorana discrimination is possible, even when neutrino masses are taken into account.

\subsubsection{Dirac Case}
For a Dirac neutrino, the effective chiral couplings are the sum of the Standard Model (SM) and New Physics (NP) contributions. Let us explicitly separate these components. The diagonal channel ($\alpha = \beta$) contains the SM interference, while the non-diagonal channels ($\alpha \neq \beta$) are driven exclusively by the NP vector boson $Z'$. We define the general chiral couplings as:
\begin{equation}
g_{L,R}^{D, \alpha\beta} = \left( g_{V}^{\text{SM}, \alpha\beta} + g_{V}^{\text{NP}, \alpha\beta} \right) \pm \left( g_{A}^{\text{SM}, \alpha\beta} + g_{A}^{\text{NP}, \alpha\beta} \right) \text{,}
\end{equation}
where $g_{V}^{\text{NP}, \alpha\beta} \equiv g_{\alpha\beta}$ is the complex vector coupling defined in Section II. Since $g_{V}^{\text{NP}, \alpha\beta}$ and $g_{A}^{\text{NP}, \alpha\beta}$ can contain both real and imaginary components, generally $|g_L^{D, \alpha\beta}| \neq |g_R^{D, \alpha\beta}|$. When summed over all final states $\beta$, the dominant diagonal SM interference term yields an asymmetric spectral shape that is highly sensitive to the real components of the couplings.

\subsubsection{Majorana Case}
For a Majorana neutrino, the inclusive sum drastically alters the expected signal due to the Hermiticity constraints established in Section II. The effective chiral couplings heavily depend on whether the transition is flavor-conserving or flavor-changing. For the diagonal channel ($\alpha = \beta$), the vector coupling identically vanishes ($g_{V}^{\text{NP}, \alpha\alpha} = 0$), strictly canceling the SM vector contribution in accordance with the PDMCT.

When deriving the scattering amplitude using Denner's rules \cite{Denner1992}, the identical nature of the Majorana fields yields a symmetry factor of $2$. We incorporate a general NP axial coupling $g_{A}^{\text{NP}, \alpha\beta}$ (which, by similar Hermiticity arguments, must be real). The chiral couplings for FCNC transitions become:
\begin{align}
g_L^{M, \alpha\beta} &= 2 \left( g_{V}^{\text{NP}, \alpha\beta} + g_{A}^{\text{NP}, \alpha\beta} \right) \text{,} \\
g_R^{M, \alpha\beta} &= 2 \left( g_{V}^{\text{NP}, \alpha\beta} - g_{A}^{\text{NP}, \alpha\beta} \right) \text{,}
\end{align}
where the vector coupling is restricted to $g_{V}^{\text{NP}, \alpha\beta} = c_{\alpha\beta} = i \operatorname{Im}(g_{\alpha\beta})$. Because the vector coupling is purely imaginary and the axial coupling is strictly real, the magnitudes of the chiral couplings are identical for FCNC transitions:
\begin{equation}
|g_L^{M, \alpha\beta}|^2 = |g_R^{M, \alpha\beta}|^2 = 4 \left[ \left(\operatorname{Im}(g_{\alpha\beta})\right)^2 + \left(g_{A}^{\text{NP}, \alpha\beta}\right)^2 \right] \text{.}
\end{equation}

Therefore, neglecting the mass-suppressed $\mathcal{O}(m_e/E_\nu)$ interference term, the total observed differential cross-section for a Majorana neutrino is the sum of a purely axial diagonal term and the FCNC terms:
\begin{equation}
\left(\frac{d\sigma_{\text{obs}}}{dT}\right)_{\text{Maj}} \propto \sigma_{\text{Axial}}^{\alpha\alpha} + \sum_{\beta \neq \alpha} \left[ \left(\operatorname{Im}(g_{\alpha\beta})\right)^2 + \left(g_{A}^{\text{NP}, \alpha\beta}\right)^2 \right] \left[ 1 + \left(1 - \frac{T}{E_\nu}\right)^2 \right] \text{.}
\end{equation}
The FCNC contribution factorizes into a symmetric shape function. This predicts a distinct, symmetric spectral distortion compared to the general Dirac case.

\subsection{The Inclusive Nature of CE$\nu$NS and FCNC}
Coherent Elastic Neutrino-Nucleus Scattering is a Neutral Current (NC) process, defined by the reaction $\nu_\alpha + A \to \nu_\beta + A$, where the nucleus remains in its ground state \cite{Freedman:1973yd}. A key experimental feature of CE$\nu$NS detectors (such as those using Liquid Argon) is that they measure only the nuclear recoil energy. They are completely blind to the flavor of the outgoing neutrino. Therefore, the total measurable cross-section is the inclusive sum over all possible final neutrino states:
\begin{equation}
\sigma_{\text{total}} = \sigma(\nu_e \to \nu_e) + \sigma(\nu_e \to \nu_\mu) + \sigma(\nu_e \to \nu_\tau) \text{.}
\end{equation}
In the absence of the FCNC $Z^{'}$, CE$\nu$NS is governed solely by the SM neutral current, which couples only to left-handed neutrinos and is flavor-diagonal. In this case, and under the same assumptions as those of the PDMCT, the coherent elastic $\nu$–nucleus cross section has the same value for Dirac and Majorana neutrinos, because the effective vector and axial couplings that enter the nuclear current are identical in both cases. Therefore, within the SM (or any model satisfying the PDMCT hypotheses), CE$\nu$NS measurements cannot distinguish the neutrino nature.

The main new effect we study appears only when a neutral vector boson $Z^{'}$ with CP-violating FCNC couplings is present. In this case, the Majorana condition forces the flavor-diagonal vector current to vanish and forbids real FCNC vector couplings, so that Majorana neutrinos couple only through imaginary, CP-violating FCNC interactions. This generates a qualitatively different dependence of the inclusive CE$\nu$NS cross section on the FCNC parameters for Dirac and Majorana neutrinos, while preserving the identical SM limit required by the PDMCT.

\subsection{Nuclear Selection Rules and Spin-Zero Targets}
When considering neutrino scattering off nuclei, an interaction involving purely left-handed neutrinos generates an effective four-fermion operator that contains both vector and axial-vector nucleon current components \cite{Denton:2022pxt}. The overall vectorial or axial-vectorial character of the measurable scattering process is fundamentally determined by the nucleon current.

To isolate the vectorial neutrino interactions, it is necessary to suppress the axial-vector contribution to the scattering cross-section. Scattering off nuclei with non-zero spin ($J \neq 0$) introduces an irreducible axial background. To isolate the vector nucleon current—and thereby directly probe the vector neutrino current—it is necessary to utilize target nuclei with zero ground-state spin, such as $^{40}\text{Ar}$, $^{76}\text{Ge}$, or $^{28}\text{Si}$ \cite{Donnelly:1979ez,Bednyakov:2018mjd}. For these targets, nuclear selection rules dictate that the axial-vector nucleon form factor is highly suppressed \cite{Akimov:2015nza,AristizabalSierra:2019zmy,Walecka:2004}. Consequently, an experiment using a spin-zero target like Liquid Argon filters the interaction, leaving only the vector part of the nucleon current and providing the ideal setup to observe the effects mandated by the PDMCT.

\subsection{The CP-Filter Mechanism at Work}

Combining the FCNC contributions to the CE$\nu$NS measurement with the spin-zero target filtration, we evaluate the distinct signatures for Dirac and Majorana neutrinos interacting via a $Z'$ boson.  Because the spin-zero nucleus (e.g., $^{40}\text{Ar}$) strictly filters out the axial-vector nucleon current \cite{Denton:2022pxt}, the observable cross-section is driven entirely by the vectorial interactions.

\begin{itemize}

\item \textbf{Dirac Neutrino:} A Dirac neutrino undergoes both diagonal (Standard Model + NP interference) and non-diagonal transitions. The cross-section is dominated by the real components of the effective couplings:\begin{equation}\sigma_{\text{Dirac}} \propto \left| Q_{\text{SM}} + i Q_{\text{NP}}^{\text{diag}} \right|^2 + \sum_{\beta \neq e} |g_{e\beta}|^2 \simeq Q_{\text{SM}}^2 + \mathcal{O}(\epsilon^2).\end{equation}where $\epsilon$ parameterizes the no-diagonal NP coupling strength.

\item \textbf{Majorana Neutrino:} As demonstrated in Section II, the vector part of the neutrino current vanishes identically for the diagonal channel ($\nu_e \to \nu_e$) due to Fermi statistics \cite{Denner1992}.
However, the doubled axial-vector part of the neutrino current interacting with the vector quark current precisely reproduces the Standard Model cross-section $Q_{\text{SM}}^2$. The New Physics flavor-changing channels ($\nu_e \to \nu_{\mu,\tau}$) survive in the presence of a purely imaginary coupling.
\begin{equation}
    \sigma_{\text{Majorana}} \propto Q_{\text{SM}}^2 + \sum_{\beta \neq e} \left|
2 c_{e\beta} \right|^2  \simeq Q_{\text{SM}}^2 + \mathcal{O}(\epsilon^2).
\end{equation}


\end{itemize}
As shown in our formalism, this leads to a drastic suppression of the New Physics deviation for the Majorana cross-section compared to the Dirac case (which retains an $\mathcal{O}(\epsilon)$ SM-NP interference term), provided $\epsilon \ll 1$.

\section{Discussion and Conclusions}

We have explored the phenomenological consequences of the ``Practical Dirac-Majorana Confusion Theorem'' (PDMCT) when applied to extensions of the Standard Model involving Flavor-Changing Neutral Currents (FCNC). As the mathematical structure of Majorana fields—dictated by Fermi-Dirac statistics and Hermiticity—forbids CP-conserving (real) vector FCNCs and flavor-diagonal vector currents, any observable vector interaction for a Majorana neutrino must arise from CP-violating FCNC couplings. Here we focus on scattering processes, but clearly, these Majorana properties could have important effects on any CP-violating observables involving Majorana neutrino interactions.

\subsection{Experimental Implications}
Focusing on scattering experiments, the results obtained can be summarized as:

\begin{itemize}
\item \textbf{CE$\nu$NS Experiments (COHERENT):} Experiments measuring Coherent Elastic Neutrino-Nucleus Scattering provide important new tools to study neutrino interactions. As derived in Section III, the effective four-fermion interaction yields both vector and axial-vector nucleon currents \cite{Denton:2022pxt}. Because spin-zero targets filter out the axial-vector nucleon component, the observable cross-section is driven strictly by the vector neutrino current. The PDMCT therefore dictates that Majorana neutrinos decouple from the diagonal vectorial $Z'$ interaction. If future high-precision measurements at COHERENT \cite{Akimov:2015nza,Akimov:2020pdx,Akimov:2017ade} deviate linearly from the Standard Model prediction (indicating an $\mathcal{O}(\epsilon)$ interference term), it would favor the Dirac nature of the neutrino. If the event rate strictly matches the SM—or tightly constrains the excess to a quadratically suppressed $\mathcal{O}(\epsilon^2)$—it would support the Majorana hypothesis, where the diagonal vectorial current is naturally filtered out.

\item \textbf{Long-Baseline Experiments (DUNE):} The Near Detector of DUNE \cite{DUNE:2020fgq,DUNE:2021tad,DeRomeri:2019kic} will collect high-statistics samples of neutrino-electron scattering events. Because the outgoing neutrino flavor is not detected, DUNE will measure the inclusive sum of all final states ($\nu_\alpha e \to \nu_x e$). Our analysis implies that DUNE could distinguish the two hypotheses by analyzing the recoil electron energy distribution. Because the PDMCT eliminates the diagonal vectorial current for Majorana states, it predicts a distinct, symmetric spectral distribution driven purely by CP-violating FCNC interactions. This strongly contrasts with the asymmetric SM-interference expected for Dirac fermions.
\end{itemize}

\subsection{Compatibility with COHERENT Data}
The recent observation of CE$\nu$NS on Argon by the COHERENT collaboration \cite{Akimov:2020pdx} measures the cross-section with a precision of approximately $\sim 30\%$. While this measurement is compatible with the Standard Model prediction, the current level of statistical uncertainty makes it impossible to directly discriminate between an $\mathcal{O}(\epsilon)$ deviation (Dirac) and an $\mathcal{O}(\epsilon^2)$ deviation (Majorana) arising from a New Physics vector boson $Z'$.

Far from ruling out our New Physics model, this highlights the critical necessity for next-generation, high-precision CE$\nu$NS detectors. If neutrinos were Dirac fermions, the interference between the Standard Model $Z$ boson and the New Physics $Z'$ vectorial current would generate a significant, linear deviation in the event rate once higher precision is achieved. The Majorana nature, in strict obedience to the PDMCT, dictates that the diagonal vectorial coupling vanishes identically. The New Physics contribution survives only through flavor-changing transitions, rendering its effect proportional to $\epsilon^2$. Therefore, bounds from future, precision COHERENT data will be profoundly sensitive to the fundamental neutrino nature.

\subsection{Conclusion}
We have presented an analysis on how the distinction between Majorana and Dirac neutrinos could be observed in scattering experiments for a new physics model based on introducing a New Physics vector boson $Z'$ with CP-violating, flavor-changing couplings. We have shown that the Majorana condition, enforced by the PDMCT, acts as a strict filter for vectorial interactions:

\begin{itemize}
\item A \textbf{Dirac neutrino} interacts via the full interference term between the Standard Model and the New Physics vector current in the flavor-diagonal channel, in addition to any non-diagonal FCNC.
\item A \textbf{Majorana neutrino} identically decouples from the flavor-diagonal vector current due to Fermi-Dirac statistics, exactly as the PDMCT mandates. Consequently, it can only interact vectorially via the CP-violating imaginary coupling in flavor-changing transitions.
\end{itemize}

We stress that, if only the SM interaction is present, our formalism reproduces the well-established result that $\nu$–e scattering and CE$\nu$NS cross sections are identical for Dirac and Majorana neutrinos. All the discriminating power we describe originates from CP-violating FCNC interactions of a $Z^{'}$ boson, which lie beyond the hypotheses of the PDMCT. 
\begin{acknowledgments}
We acknowledge financial support from SNI (M\'exico). A.Y. and D.D. are grateful to Dr. Juan Barranco for useful comments on this work.
\end{acknowledgments}

\end{document}